\documentclass[twocolumn,prl,showpacs,preprintnumbers,amsmath,amssymb]{revtex4}
\usepackage{graphicx}% Include figure files
\usepackage{dcolumn}% Align table columns on decimal point
\usepackage{bm}% bold math

\begin{document}
\preprint{Xu $et$  $al$., submitted to PRL}

\title{Band dependent normal-state coherence in Sr$_{2}$RuO$_{4}$: Evidence from Nernst effect and thermopower measurements}

\author {X. F. Xu,$^{1}$ Z. A. Xu,$^{1,}$\footnote[1]{zhuan@zju.edu.cn} T. J. Liu,$^{2}$ D. Fobes,$^{2}$ Z. Q. Mao,$^{2}$ J. L. Luo,$^{3}$
Y. Liu$^{4,1,}$\footnote[2]{liu@phys.psu.edu}}

\affiliation{$^{1}$Department of Physics, Zhejiang University,
Hangzhou 310027, China}

\affiliation{$^{2}$Department of Physics, Tulane University, New
Orleans, LA 70118, U.S.A.}

\affiliation{$^{3}$Institute of Physics, Chinese Academy of
Sciences, Beijing 100080, China.}

\affiliation{$^{4}$Department of Physics and Material Research
Institute, The Pennsylvania State University, University Park, PA
16802, U.S.A.}

\date{\today}% It is always \today, today,

\begin{abstract}
We present the first measurement on Nernst effect in the normal
state of odd-parity, spin-triplet superconductor
Sr$_{2}$RuO$_{4}$. Below 100 K, the Nernst signal was found to be
negative, large, and, as a function of magnetic field, nonlinear.
Its magnitude increases with the decreasing temperature until
reaching a maximum around $T^*$ $\approx$ 20 - 25 K, below which
it starts to decrease linearly as a function of temperature. The
large value of the Nernst signal appears to be related to the
multiband nature of the normal state and the nonlinearity to
band-dependent magnetic fluctuation in Sr$_{2}$RuO$_{4}$. We argue
that the sharp decrease in Nernst signal below $T^*$ is due to the
suppression of quasiparticle scattering and the emergence of
band-dependent coherence in the normal state. The observation of a
sharp kink in the temperature dependent thermopower around $T^*$
and a sharp drop of Hall angle at low temperatures provide
additional support to this picture.

\end{abstract}


\maketitle

Support for an odd-parity, spin-triplet pairing in
Sr$_{2}$RuO$_{4}$ by increasing number of
experiments\cite{MM03,Liu04,AK06,vH06}, including in particular
the first phase-sensitive measurements\cite{Liu04}, has provided
strong motivation to pursue a detailed understanding of the
mechanism of superconductivity in Sr$_{2}$RuO$_{4}$. Rice and
Sigrist\cite{RS95} suggested that ferromagnetic (FM) fluctuation
might be responsible for spin-triplet superconductivity in
Sr$_{2}$RuO$_{4}$. Inelastic neutron scattering (INS)
measurements\cite{INS} revealed magnetic fluctuation with peaks in
the dynamic susceptibility only around $k$-vectors
($\pm$0.6,$\pm$0.6, 0)($\pi$/$a$), where $a$ is the lattice
constant, suggesting that incommensurate magnetic fluctuation
(IMF) dominates in Sr$_{2}$RuO$_{4}$. Theoretical
calculations\cite{Mazin99} suggest that the IMF originates from
the one-dimensional (1D) $d_{xz,yz}$ bands in Sr$_{2}$RuO$_{4}$,
and spin-singlet rather than spin-triplet superconductivity is
favored in these 1D bands. These observations have raised
questions on the mechanism on superconductivity in
Sr$_{2}$RuO$_{4}$. So far, models based on FM\cite{FM},
antiferromagnetic (AF) fluctuations\cite{AFM}, spin-orbital
coupling\cite{SOCoupling}, or Hund's rule coupling\cite{Hund} for
superconductivity in Sr$_{2}$RuO$_{4}$ have been proposed. But
none has gained universal acceptance.

The Fermi surface of Sr$_{2}$RuO$_{4}$ consists of three
cylindrical sheets\cite{Multibanda,Multibandb}, one hole (the
$\alpha$) and two electron-like (the $\beta$ and the $\gamma$)
bands. It was suggested that superconductivity in
Sr$_{2}$RuO$_{4}$ is band-dependent with its energy gap large on
the $\gamma$ band but tiny on the $\alpha$ and the $\beta$
bands\cite{ARS97}. The interesting question is whether the
orbital-dependent superconductivity comes from orbital-dependent
normal-state properties to begin with. This idea can actually be
traced back to the original paper of Baskaran\cite{Hund}, who was
among the first to suggest that Sr$_{2}$RuO$_{4}$ is a $p$-wave
superconductor. Thermoelectric measurements provide a natural
probe for the orbital-dependent physical properties, as
demonstrated previously\cite{NbSe2}. In this Letter, we present
the first comprehensive study of the Nernst effect and thermopower
in Sr$_{2}$RuO$_{4}$. Our results show clearly that quasiparticle
scattering is strongly suppressed and band-dependent coherence
emerges at low temperatures.

\begin{figure}[tbp]
\includegraphics[width=8cm]{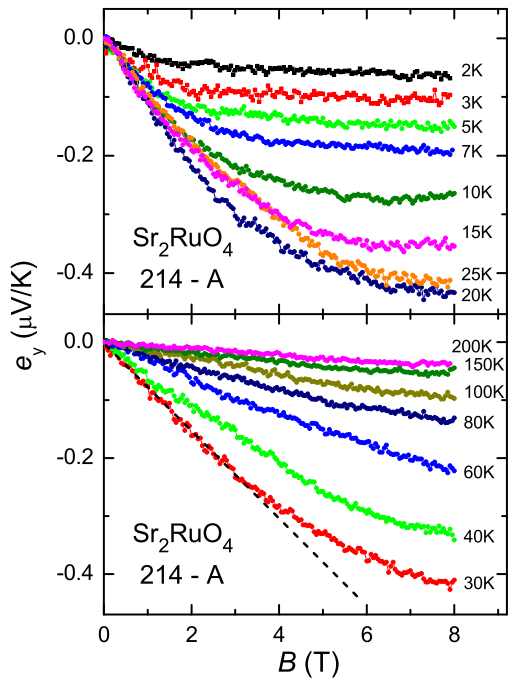}
\caption{(Color Online) Magnetic field ($B$) dependence of the
Nernst signal ($e_{y}$) at different temperatures for Sample
214-A. The linear part of $e_{y}$ is indicated by a dash line for
30K.}
\end{figure}

Single crystals of Sr$_{2}$RuO$_{4}$ were grown by the
floating-zone method as reported previously\cite{Mao99}. Three
single crystals, labelled as 214-A, B and C, were measured in this
experiment. The values of the in-plane residual resistivity were
found to be less than 0.5 $\mu$$\Omega$cm for all three crystals,
confirming the ultra-high quality of the crystals\cite{Mao99}. In
addition, we measured $each$ of these crystals in a Quantum Design
MPMS-5 system to ensure the absence of the intergrowth of other
phases in the Ruddlesden-Popper series of
Sr$_{n+1}$Ru$_n$O$_{3n+1}$. The thermoelectric properties were
measured by the steady-state technique. The magnetic field was
applied along the $c$ axis. The temperature gradient, around 0.5
K/mm, was applied in the $ab$ plane and determined by a pair of
differential Type E thermocouples. All measurements were performed
in a Quantum Design PPMS-9 system.

Traces of Nernst signal as a function of magnetic field for Sample
214-A at various temperatures are displayed in Fig. 1. Above 100
K, the Nernst signal is small and linear as a function of magnetic
field. As temperature decreases, the magnitude of the Nernst
signal increases at a given field, reaching a large value of about
-0.43 $\mu$V/K under $B$ = 8 T at 20 K. However, the growth of the
magnitude of the Nernst signal, $|e_{y}|$, is reversed around
$T$*, below which $|e_{y}|$ decreases linearly. An equally
striking feature is that, as the temperature is lowered, the
Nernst signal is found to become nonlinear above a characteristic
field. The "characteristic" field decreases with decreasing
temperature, indicating that the nonlinearity becomes increasingly
pronounced at low temperatures. Similar data are also obtained for
the other two samples.

\begin{figure}[tbp]
\includegraphics[width=8cm]{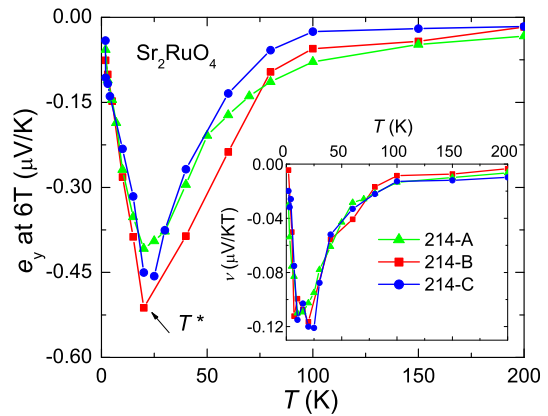}
\caption{(Color Online) Temperature ($T$) dependence of the Nernst
signal ($e_{y}$) at $B$ = 6 T for the three measured samples.
Inset: the temperature dependence of the Nernst coefficient,
$\nu$, determined from the initial slope of the $e_{y}$ vs. $B$
curves for the same samples.}
\end{figure}

Figure 2 shows the temperature dependence of Nernst signal at a
magnetic field of 6 T obtained from all three samples measured.
The magnitude of Nernst signal starts to increase around 100 K as
the temperature is lowered. After reaching a maximum around $T^*$,
the magnitude of the Nernst signal is seen to decrease linearly
with the decreasing temperature, approaching zero as temperature
decreases to zero. The temperature dependence of the Nernst
coefficient, $\nu$ $\equiv$ d$e$$_{y}$/d$B$ determined from the
initial slope of the $e$$_{y}$ $vs$. $B$ curves, shows similar
behavior (the inset of Fig. 2).

A large and strongly nonlinear Nernst signal observed in a
metallic sample is highly unusual. In a single-band conventional
metal, the Nernst signal is usually small because of the
Sondheimer cancellation\cite{Wang01}, and linearly dependent on
magnetic field. We are aware of only few examples showing
otherwise. The first is the vortex motion in the mixed state of a
type II superconductor, as observed in the mixed state and a
certain temperature range above $T_{c}$ of high-$T_{c}$
superconducting cuprates\cite{XuNature, Wang01, Wang06}; The
second example is that a large and nonlinear Nernst signal was
found below and even above the Curie temperature in a
ferromagnet\cite{Lee04}; The third example was found in the Kondo
lattice, heavy Fermion superconductor CeCoIn$_{5}$\cite{CeCoIn5},
even though its physical origin is not understood; The fourth
example is related to the difference in scattering rates of
different energy bands in a multiband metal such as
NbSe$_{2}$\cite{NbSe2}. The presence of two types of carriers,
invalidates the Sondheimer cancellation, resulting in a large
Nernst signal.

For Sr$_{2}$RuO$_{4}$, no superconducting fluctuation is expected
at a temperature as large as 100 K. Furthermore, the Nernst signal
is negative rather than positive as it would be expected in a type
II superconductor. On the other hand, Sr$_{2}$RuO$_{4}$ is a
multiband metal whose Nernst signal is enhanced as seen in
NbSe$_2$\cite{NbSe2}. In a simple two-band model, the Nernst
signal can be express as
\begin{equation}
e_{y}=S(\frac{\alpha^{h}_{xy}+\alpha^{e}_{xy}}{\alpha^{h}_{xx}+\alpha^{e}_{xx}}-\frac{\sigma^{h}_{xy}+\sigma^{e}_{xy}}{\sigma^{h}_{xx}+\sigma^{e}_{xx}}),
\end{equation}
where $S$ is the thermopower, $\alpha$ is the Peltier conductivity
tensor and $\sigma$ is the electric conductivity tensor with the
superscripts e and h referring to electrons and holes. The
subscripts $xx$ and $xy$ refer to the diagonal and off-diagonal
components of the tensors. It can be seen that the Nernst signal
is sensitive to the different temperature dependence of scattering
rates in different bands.

The observed nonlinearity in the Nernst signal, on the other hand,
appears to be related to magnetic fluctuation. As pointed above,
IMF originating from $\alpha$ and $\beta$ bands dominates in
Sr$_{2}$RuO$_{4}$\cite{INS, NMR}. If the IMF is suppressed so that
its contribution to the Nernst signal decreases at high magnetic
fields, or alternatively, FM fluctuation is induced by an applied
field, a nonlinear Nernst signal can be understood. In other case,
the maximum in the temperature dependence of the Nernst signal
near $T^*$ (Fig. 2) suggests a sharp change in band-dependent
quasiparticle scattering rates or the density of states at this
temperature. To address the latter possibility, we measured the
specific heat and found no anomaly present around $T^*$ (Inset of
Fig. 3). Our results suggest that it is unlikely that the density
of states is changed around $T^*$. On the other hand, below $T^*$,
the Fermi liquid behavior, $i$.$e$, the $T^{2}$ behavior, was
found in both in- and out-of-plane resistivities. This is
consistent with linear temperature dependence of Nernst signal or
Nernst coefficient observed in Sr$_{2}$RuO$_{4}$ below $T^*$
\cite{Oganesyan04}. The emergence of the Fermi liquid and the
change in temperature dependence of the Nernst signal at the same
temperature therefore signal an important change in the character
of the normal state. We argue that a coherence state emerges at $T
<$ $T^*$, which will be further discussed below.

Since thermopower is sensitive to the change of quasiparticle
scattering rates in different bands, any change in the electronic
state should also result in a sharp feature in the temperature
dependence of the thermopower. As shown in the upper panel of Fig.
3, the $S$($T$) curves indeed exhibit a sharp change of slope
around $T^*$. This change becomes even more striking in the plot
of d$S$/d$T$ $vs$. $T$ as shown in the lower panel of Fig. 3. Such
a feature in thermopower was found previously in
Sr$_{2}$RuO$_{4}$\cite{seebeck}, although not as pronounced as
seen in the present work.

The Hall coefficient was found previously to change its sign from
hole-like (suggesting that the $\alpha$ band dominates) above
$T^*$, to electron-like below this
temperature\cite{Shirakawa95,Liu99,Mackenzie96}. It is known that
Hall angle, rather than Hall coefficient, is directly related to
the scattering rate of the quasiparticle scattering. In Fig. 4, we
show the temperature dependence of Hall angle, which shows a steep
drop at low temperatures. This drop marks the increase of the
scattering time, which may be taken as direct evidence of the
existence of normal-state coherence. The temperature dependence of
Hall coefficient can be well understood\cite{Mackenzie96} in a
multi-band model\cite{ong}. The sharp drop of $R_H$, thus
tan$\theta$, below 25 K, indicates a decreasing $l_h$/$l_e$ ratio,
where $l_h$($l_e$) is the mean free path in the hole-like
(electron-like) band. The sharp increase in $l_e$ should result
from the band-dependent change in the scattering time. As pointed
out before, quasiparticle scattering in $\alpha$ as well as
$\beta$ band is dominated by IMF. However, no change was found in
IMF around $T^*$\cite{INS02}. Therefore it is reasonable to
conclude that the change in quasiparticle scattering rate revealed
by the Nernst effect and Hall effect measurements must be limited
to the $\gamma$ band. The increase in $l_e$ is consistent with the
emergence of a coherence state in the electron-like $\gamma$ band.
This implies that the coherence among quasiparticles occurs in
this particular band originating from $d_{xy}$ orbitals.

\begin{figure}[tbp]
\includegraphics[width=8cm]{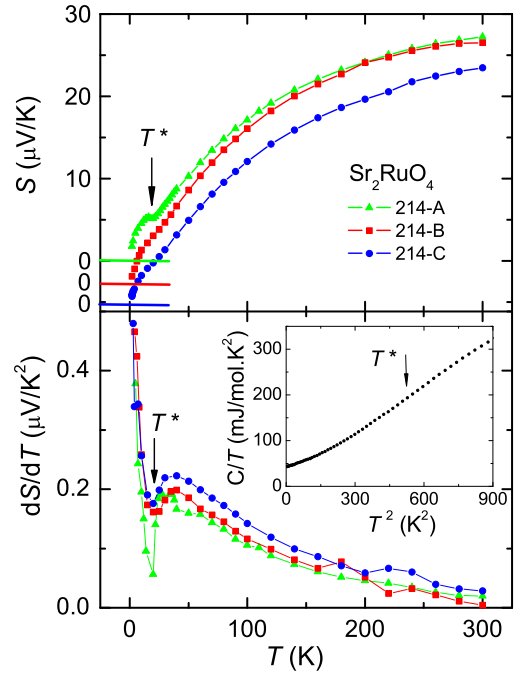}
\caption{(Color Online)  Upper panel: Temperature dependence of
the thermopower. The data of the samples 214-B and 214-C are
shifted -3 $\mu$V/K and -6 $\mu$V/K for clarity. Lower panel: The
derivative of thermopower, d$S$/d$T$, $vs$. $T$. Inset: $C$/$T$
$vs$. $T^{2}$.}
\end{figure}

\begin{figure}[tbp]
\includegraphics[width=8cm]{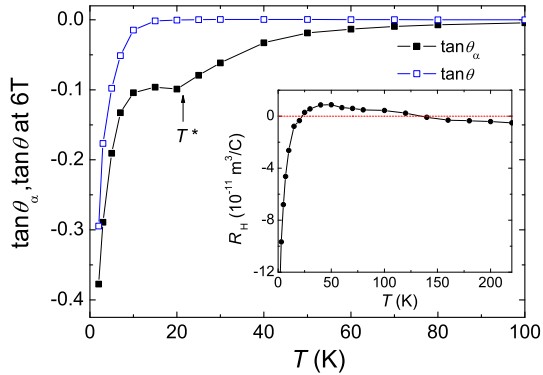}
\caption{(Color Online)  Temperature dependence of Hall angle and
"Peltier Hall angle" for samples 214-A taken at $B$ = 6 T. The
inset shows the temperature dependence of the Hall coefficient.}
\end{figure}

Nernst measurement probes both the off-diagonal Peltier current
and ordinary Hall current characterized by
\begin{equation}
e_{y}=\rho\alpha_{xy}+\rho_{xy}\alpha_{xx}
=S(\tan\theta_{\alpha}-\tan\theta)
\end{equation}
where the "Peltier Hall angle" tan$\theta_{\alpha}$ is defined as
tan$\theta_{\alpha}$=$\alpha_{xy}$/$\alpha_{xx}$. Actually each
angle includes the contributions from both electron and hole bands
(see Eq. 1). The "Peltier Hall angle" tan$\theta_{\alpha}$ is also
shown in Fig. 4. At temperatures above 25 K, the Hall angle
tan$\theta$ shows very weak temperature dependence, and the
absolute magnitude of the "Peltier Hall angle" increases
gradually. It is clear that the gradual increase in $e_y$
originates from the increase of tan$\theta_{\alpha}$. Around
$T^*$, a kink in tan$\theta_{\alpha}$ was observed. While the Hall
angle probes the scattering time, the "Peltier Hall angle" is
sensitive to the energy dependence of the scattering
time\cite{Wang01}. The band dependent coherence state is
responsible for features seen in both tan$\theta$ and
tan$\theta_{\alpha}$.

An interesting question is whether the quasiparticle coherence in
the $\gamma$ band below $T^*$ can be viewed by the opening of a
pseudogap. Infrared spectroscopy measurements did find evidence
for a gap of 6.3 meV opening in the normal state of
Sr$_{2}$RuO$_{4}$\cite{Infrared}. A kink in the temperature
dependence of 1/$T$$_{1}$ around 70 K in the NMR measurements is
consistent with the opening of a pseudogap\cite{NMR}. However, the
presence of Fermi liquid behavior in Sr$_{2}$RuO$_{4}$ seems to be
inconsistent with the pseudogap idea\cite{Liu99}. The new insight
provided by the present study is that the pseudogap may be
band-dependent, and coexist with the Fermi liquid behavior.

The emergence of normal-state coherence should be related to
superconductivity in Sr$_{2}$RuO$_{4}$. It was proposed previously
\cite{ARS97} that the $\gamma$ band is the active band that gives
rise to spin-triplet superconductivity in Sr$_{2}$RuO$_{4}$. It
seems natural that the coherence among quasiparticles and
superconductivity in the $\gamma$ band are correlated.
Alternatively, if the coherence below $T^*$ reflects a hidden
order of a non-superconducting state, it may actually compete with
rather than help superconductivity. In this regard, it may be
useful to point out that in Sr$_{3}$Ru$_{2}$O$_{7}$, a
paramagnetic compound closely related to Sr$_{2}$RuO$_{4}$,
tendencies for both FM and AF orderings were found to coexist, and
evidently compete for stability\cite{Liu01}. Thus our
thermoelectric measurements may provide insight into the question
raised originally by Baskaran\cite{Hund} - why is the $T_{c}$ of
Sr$_{2}$RuO$_{4}$ so low?

\begin{acknowledgments}
The authors wish to thank D. F. Agterberg, E. A. Kim, M. Sigrist,
C. Wu, and F. C. Zhang for helpful discussions. The work is
supported by the NSFC (Grant Nos 10634030 and 10628408) and PCSIRT
(IRT0754) at Zhejiang University, by DOE under Grant
DE-FG02-04ER46159 as well as by DOD ARO under Grant
W911NF-07-1-0182 at Penn State, and by Research Corporation and
NSF under grant DMR-0645305 and DOE under grant DE-FG02-07ER46358
at Tulane.
\end{acknowledgments}

\end{document}